\documentclass[referee]{aa}

\usepackage{graphicx}
\usepackage{color}

\begin{document}

\title{Antiprotons from primordial black holes}

\author{Aur\'elien Barrau\inst{1,3}, Ga\"elle Boudoul\inst{1,3},
Fiorenza Donato\inst{2}, David Maurin\inst{2},\\
Pierre Salati\inst{2,4}, Richard Taillet\inst{2,4}
}

\offprints{A. Barrau} \mail{barrau@isn.in2p3.fr}

\institute{
ISN Grenoble, 53 av des Martyrs, 38026 Grenoble cedex, France
\and
LAPTH, B.P. 110 Annecy-le-Vieux, 74941, France
\and
Universit\'e Joseph Fourier, Grenoble, 38000, France.
\and
Universit\'e de Savoie, Chamb\'ery, 73011, France.
}

\date{Received;Accepted}

\titlerunning{Antiprotons from primordial black holes}

\authorrunning{Barrau et al.}

\abstract{
Primordial black holes ({\sc pbh}s) have motivated many studies since it was
shown that they
should evaporate and produce all kinds of particles (Hawking
\cite{Hawking1}).
Recent experimental measurements of cosmic rays with great accuracy,
theoretical
investigations on the possible formation mechanisms and detailed
evaporation processes
have revived the interest in such astrophysical objects. This article
aims at using
the latest developments on antiproton propagation models (Maurin et
al. \cite{David} and
Donato et al. \cite{Fiorenza}) together with new data from BESS,
CAPRICE and AMS experiments
to constrain the local amount of {\sc pbh} dark matter.
Depending on the diffusion halo parameters and on the details of
emission mechanisms,
we derive an average upper limit of the order of
$\rho_{\odot}^{PBH}\approx 1.7\cdot10^{-33}$g cm$^{-3}$.
}

\maketitle

\section*{Introduction}

Very small black holes could have formed in the early universe from
initial density
inhomogeneities (Hawking \cite{Hawking2}), from phase transition
(Hawking \cite{Hawking3}),
from collapse of cosmic strings (Hawking \cite{Hawking4}) or as a
result of a softening of the
equation of state (Canuto \cite{Canuto}). It was also shown
by Choptuik (\cite{Choptuik}) and,
more recently, studied in the framework of double inflation (Kim
\cite{Kim}), that {\sc pbh}s could even
have formed by near-critical collapse in the expanding universe.

The interest in primordial black holes was clearly revived in the
last years for several reasons.
First, new experimental data on gamma-rays (Connaughton
\cite{Connaughton}) and cosmic rays (Maki
{\it et al.} \cite{Orito}) together with the construction of
neutrinos detectors (Bugaev \& Konishchev
\cite{Bugaev}), of extremely high-energy particle observatories
(Barrau \cite{Barrau})
and of gravitational waves
interferometers (Nakamura {\it et al.} \cite{Nakamura}) give
interesting investigation means to look for indirect signatures of
{\sc pbh}s. Then, primordial black holes have
been used to derive a quite stringent limit on the spectral index of
the initial scalar perturbation
power spectrum due to a dust-like phase of the cosmological expansion
(Kotok \& Naselsky \cite{Kotok}).
It was also found that {\sc pbh}s are a great probe of the
early universe with a varying
gravitational constant (Carr \cite {Carr2}).
Finally, significant progress have been made in the understanding of
the evaporation
mechanism itself, both at usual energies (Parikh \& Wilczek
\cite{Parikh}) and in the near-planckian
tail of the spectrum (Barrau \& Alexeyev \cite{Barrau2}, Alexeyev {\it et
al.}
\cite{Stas}).

Among other cosmic rays, antiprotons are especially interesting as
their secondary flux is both rather small
(the $\bar{p}/p$ ratio is lower than $10^{-4}$ at all energies) and
quite well known.
This study aims at deriving new estimations of the {\sc pbh} local density,
taking into account the
latest measurements of the antiproton spectrum and realistic models
for cosmic ray propagation.
Contrarily to what has been thought in early studies on antiproton
spectra (see, {\it e.g.},
Gaisser \& Schaefer \cite{Gaisser} for a review) and to what is still
assumed in some recent {\sc pbh} studies ({\it e.g.}
Kanazawa {\it et al.} \cite{Kanazawa}) there is no need for any
exotic astrophysical source to account for the
measured $\bar{p}$ flux. On the other hand, the very good agreement
between experimental data and theoretical
predictions with a purely secondary origin of antiprotons
(see Donato {\em et al.} \cite{Fiorenza}) allows to derive quite stringent
upper limit on the amount of {\sc pbh} dark matter. This article
focuses on this
point and is organised in the following way : the first section reminds
general consideration on the evaporation process and on the subsequent
fragmentation phenomena. The second part is devoted to the
primary antiproton sources while the third part describes all
propagation aspects, with a particular emphasis on the uncertainties
coming from astrophysical parameters (mostly the magnetic halo thickness) and
nuclear physics cross-sections. Next, the resulting ``top of
atmosphere" (TOA) antiproton
spectra for different formation conditions and for different possible
emission models are studied.
Finally, we derive the resulting
upper limits on the local {\sc pbh} density and consider possible
future developments both on the
experimental and theoretical sides.

\section{Antiproton emission from  {\sc pbh}s}

\subsection{Hawking process}

The Hawking black hole evaporation process can be intuitively
understood as a quantum creation of
particles from the vacuum by an external field (see Frolov \& Novikov 
\cite{Frolov} for
more details). This can occur as a result of
gravitation in a region where the Killing vector is spacelike,
{\it i.e.} lying inside the $\xi^2=0$ surface, which is the event
horizon in a static spacetime.
This basic argument shows that particle creation by a gravitational
field, in a stationary case, is possible only if it
contains a black hole. Although
very similar to the effect of particle creation by an electric field,
the Hawking process has a
fundamental difference: since the states of negative energy are
confined inside the hole, only one
of the created particles can appear outside and reach infinity. This
means that the classical
observer has access to only a part of the total quantum system.

To derive the accurate emission process, which mimics a Planck law,
Hawking used the usual quantum
mechanical wave equation for a collapsing object with a postcollapse
classical curved metric (Hawking \cite{Hawking5}). He found
that the emission spectrum for
particles of energy $Q$ per unit of time $t$ is, for each degree of
freedom:
\begin{displaymath}
      \frac{{\rm d}^2N}{{\rm d}Q{\rm 
d}t}=\frac{\Gamma_s}{h\left(\exp\left(\frac{Q}{h\kappa/4\pi^2c}\right)-(-1)^{2s}\right)}
\end{displaymath}
where contributions of electric potential  and angular velocity have
been neglected since the
black hole discharges and finishes its rotation much faster than it
evaporates (Gibbons \cite{Gibbons}
and Page \cite{Page1}). $\kappa$ is the surface gravity, $s$ is the
spin of the emitted species and
$\Gamma_s$ is the absorption probability. If we introduce the Hawking
temperature (one of the rare
physics formula using all the fundamental constants) defined by
\begin{displaymath}
      T=\frac{hc^3}{16\pi k G M}\approx\frac{10^{13}{\rm g}}{M}{\rm GeV}
\end{displaymath}
the argument of the exponent becomes simply a function of $Q/kT$.
Although the absorption probability is often approximated by its
relativistic limit,
we took into account in this work its real expression for
non-relativistic particles:
\begin{displaymath}
      \Gamma_s=\frac{4\pi \sigma_s(Q,M,\mu)}{h^2c^2}(Q^2-\mu^2)
\end{displaymath}
where $\sigma_s$ is the absorption cross-section computed numerically
(Page \cite{Page2}) and $\mu$ is the rest mass of the emitted particle.

\subsection{Hadronization}

As it was shown by MacGibbon and Webber (\cite{MacGibbon1}), when the
black hole temperature is
greater than the quantum chromodynamics confinement scale
$\Lambda_{QCD}$, quarks and gluons jets are
emitted instead of composite hadrons. To evaluate the number of
emitted antiprotons, one therefore
needs to perform the following convolution:
\begin{displaymath}
      \frac{{\rm d}^2N_{\bar{p}}}{{\rm d}E{\rm d}t}=
      \sum_j\int_{Q=E}^{\infty}\alpha_j\frac{\Gamma_j(Q,T)}{h}
      \left(e^{\frac{Q}{kT}}-(-1)^{2s_j}\right)^{-1}
      \times\frac{{\rm d}g_{j\bar{p}}(Q,E)}{{\rm d}E}{\rm d}Q
\end{displaymath}
where $\alpha_j$ is the number of degrees of freedom, $E$ is the
antiproton energy and
${\rm d}g_{j\bar{p}}(Q,E)/{\rm d}E$ is the normalized differential
fragmentation function, {\it i.e.}
the number of antiprotons between $E$ and $E+{\rm d}E$ created by a
parton jet of type $j$ and energy
$Q$ (including decay products). The fragmentation functions have been 
evaluated with the
high-energy physics frequently used event generator
{\sc pythia}/{\sc jetset} (Tj\"{o}strand \cite{Tj}),
based on the
so-called string fragmentation
model.

\section{Primary sources}

In order to compute the antiproton spectrum
for a given local {\sc pbh} density,
the number $q^{PBH}_{\bar{p}}(r,z,E)$ of  antiprotons emitted
with kinetic energy between $E$ and $E+{\rm d}E$ per unit volume and
time, must be evaluated.
It is proportional to the number ${\rm d}^2n/{\rm d}M{\rm d}V$,
of {\sc pbh}s per unit of mass and
volume and to the individual flux $d^2N_{\bar{p}}/dE\, dt$ emitted by
one {\sc pbh}, so that
\begin{displaymath}
      q^{prim}(r,z,E) = \int\frac{d^2N_{\bar{p}}(M,E)}{dE\, dt} \cdot
      \frac{d^2n(r,z)}{dM \, dV} dM
\end{displaymath}
where $r$ and $z$ are the cylindrical coordinates describing position
in our Galaxy.
As the physics of evaporation and the mass spectrum of {\sc pbh}  do
not depend on their numerical density,
the primary source term can be split into a spatial and a spectral
dependance, as
\begin{equation}
       q^{prim}(r,z,E) = q^{prim}(r,z) \times Q^{PBH}(E)
       \label{source_primaire}
\end{equation}
with
\begin{equation}
      q^{prim}(r,z) = \frac{\rho^{PBH}(r,z)}{\rho^{PBH}_\odot}
      \;\;\; \mbox{and} \;\;\;
      Q^{PHB}(E)=\int_{M_{min}}^{M_{max}} \frac{d^2N_{\bar{p}}}{dE\, dt} \cdot
      \frac{d^2n_\odot}{dM \, dV} dM
      \label{splotchhh}
\end{equation}

\subsection{Spatial distribution}
\label{distrib}

Primordial black holes should have followed the dark
matter particles during the formation of halos, so that they may have
the same spatial distribution.
Unfortunately, this distribution is not very well known, and
several independent evidences are contradictory.
In the absence of a clear answer to this problem
({\em e.g.} see Ghez. {\it et al.} \cite{Ghez}, Gondolo \& Silk
\cite{Gondolo}, Debattista and Sellwood \cite{debattista1,debattista2}),
several profiles for the {\sc pbh} distribution can be used, with the
generic form
\begin{equation}
        \frac{\rho^{PBH}(r,z)}{\rho_{\odot}^{PBH}} = \left(
        \frac{R_\odot}{\sqrt{r^2+z^2}}
        \right)^\gamma
        \left( \frac{R_c^\alpha + R_\odot^\alpha}{R_c^\alpha +
(\sqrt{r^2+z^2})^\alpha}
        \right)^\epsilon
        \label{dep_spatiale}
\end{equation}
where spherical symmetry has been assumed.
Different cases have been considered, with numerical values taken in
Calc\'aneo-Rold\'an and Moore (\cite{calcaneo}).
Numerical simulations point toward singular profiles with
$\gamma = 1.5$, $\alpha = 1.5$, $\epsilon = 1$ and $R_c=33.2
\, \mbox{kpc}$ (Moore \cite{moore}) or
$\gamma = 1$, $\alpha = 1$,  $\epsilon = 2$ and $R_c=27.7
\, \mbox{kpc}$ (Navarro, Frank \& White  \cite{NFW}, hereafter NFW).
We also considered an isothermal profile with $\gamma = 0$, $\alpha =
2$ and $\epsilon = 1$ and a modified isothermal profile with
$\gamma = 0$, $\alpha = 2$ and $\epsilon = 1.5$ and $R_c=24.3 \,
\mbox{kpc}$.

\subsection{Spectral dependence}

It can be seen from eq. (\ref{splotchhh}) that
the keypoints to derive the $Q^{PHB}(E)$ term are the shape of
the mass spectrum today and the boundaries of the integral.
The choice of $M_{min}$ and $M_{max}$ is discussed later.

\subsubsection{{\sc pbh} mass spectrum}

The mass spectrum today is the result of the evolution of the initial
mass spectrum in time. The usual
picture is based on the idea that a {\sc pbh} can form if an overdense
region collapses a contrast density $\delta$ (Harrison
\cite{Harrison}, Carr \& Hawking \cite{Carr3}) so that $1/3 \leq \delta \leq 1$.
For a scale invariant power spectrum, this leads to
\begin{displaymath}
      \left(\frac{{\rm d}n}{{\rm d}M_{init}}\right) \propto
      M_{init}^{-5/2}.
\end{displaymath}
This mass spectrum will be regarded as the {\it standard model} (see, {\it
e.g.} Bugaev \&
Konishev \cite{Bugaev} for a review). To deduce the mass spectrum
today from this
initial one, the mass loss rate must be evaluated: ${\rm d}M/{\rm d}t
\approx \{ 7.8d_{s=1/2}+3.1d_{s=1} \} \times 10^{24}$
g$^3$ s$^{-1}$ with $d_{s=1/2}=90$ and $d_{s=1}=27$ (MacGibbon \cite{MacGibbon3}).
If we approximate the
previous law with $\alpha \approx
const$, which is correct as far as the number of degrees of freedom
does not increase dramatically fast, the
mass can be written as a function of time as:
\begin{displaymath}
      M_{init}\approx(3\alpha t + M^3)^{1/3}.
\end{displaymath}
It is then straightforward to see that with
\begin{displaymath}
      \frac{{\rm d}n}{{\rm d}M}=\frac{{\rm d}n}{{\rm d}M_{init}}\cdot
      \frac{{\rm d}M_{init}}{{\rm d}M}
\end{displaymath}
the resulting spectrum today is characterised by
\begin{displaymath}
      \frac{{\rm d}n}{{\rm d}M}\propto M^2{\rm~for~M}<M_*
\end{displaymath}
\begin{displaymath}
      \frac{{\rm d}n}{{\rm d}M}\propto M^{-5/2}{\rm~for~M}>M_*
\end{displaymath}
where $M_* \approx 5\times 10^{14} {\rm g}$ is the initial mass of a
{\sc pbh} expiring nowadays.
In this model, the mass of a {\sc pbh} formed at time $t$ is determined by
the horizon mass at this
epoch,
\begin{displaymath}
      M=\frac{1}{8}\frac{M_{Pl}}{t_{Pl}}t.
\end{displaymath}
In the usual cosmological picture, only {\sc pbh}s formed after inflation
should be taken into
account as those produced before would be extremely diluted due to
the huge increase of the
cosmic scale factor. The end-time of inflation $t_{RH}$ being related to the
reheating temperature $T_{RH}$ by
\begin{displaymath}
      t_{RH}\approx0.3g^{-1/2}\frac{M_{Pl}}{T^2_{RH}}
\end{displaymath}
where $g \approx 100$ is the number of degrees of freedom in the
early universe, the minimal
mass of {\sc pbh} in the initial spectrum can be given as a function
$T_{RH}$.

It should be emphasized that the shape of the spectrum below $M_*
\approx (3\alpha t)^{1/3}$ does not depend on any assumption about the
initial mass spectrum. Whatever the initial spectrum, it should
increase as $M^2$ today: this is only due to the ${\rm
d}M_{init}/{\rm d}M$ term which
is proportional to $M^2$ for small masses whereas the ${\rm d}n/{\rm
d}M_{init}$ term is nearly constant. As shown in the next
section, the antiproton emission is governed by {\sc pbh}s with masses below
$10^{14}$g. The results derived are therefore independent of the details of
the formation mechanism.

\subsubsection{Cumulative source before propagation}
\label{cumul}

\begin{figure}
\centerline{\includegraphics*[width=\textwidth]{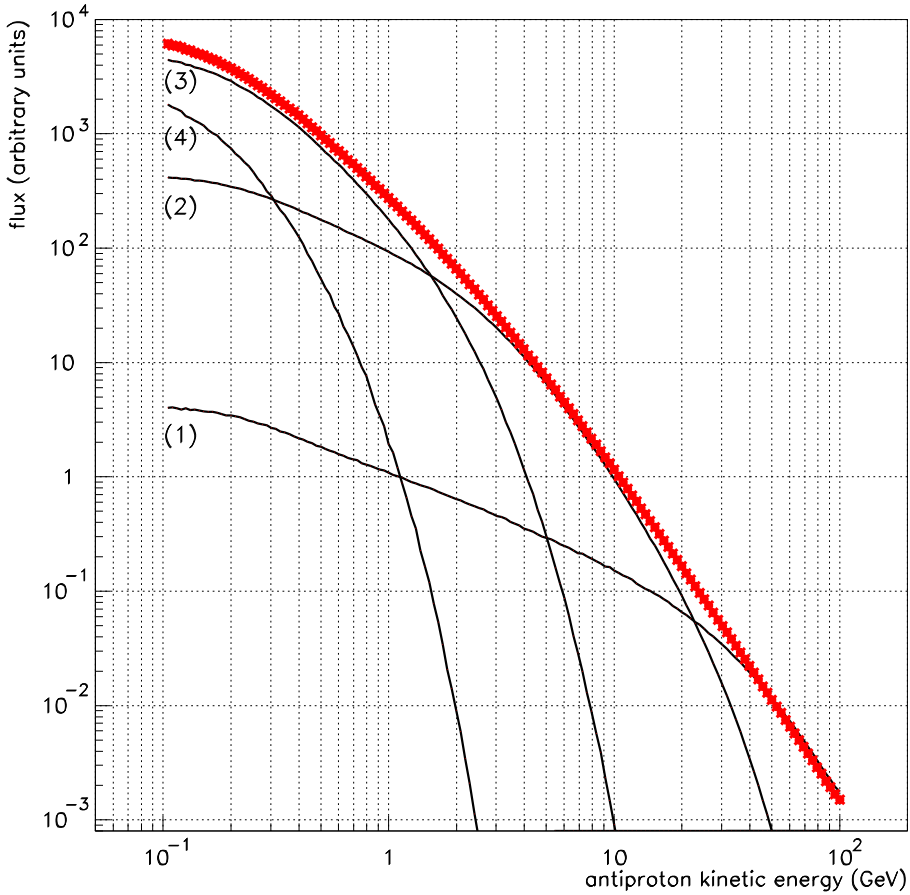}}
\caption{Primary antiproton flux with the standard mass spectrum
before propagation (in arbitrary
units). Curve (1) is for $M\in[M_{Pl},10^{12}{\rm g}]$, curve (2) is for
$M\in[10^{12}{\rm g},10^{13}{\rm g}]$, curve (3) is for
$M\in[10^{13}{\rm
g},5\cdot10^{13}{\rm g}]$, curve (4) is for $M>5\cdot 10^{13}{\rm g}$
and the thick line
is the full spectrum.}
\label{pbar}
\end{figure}

Fig.~\ref{pbar} gives the antiproton flux after convolution with the
{\sc pbh} mass spectrum before propagation. As it could be
expected, {\sc pbh}s heavier than $M_*$ nearly do not contribute as both
their number density and
their temperature is low. On the other hand, only {\sc pbh}s with very
small masses, {\it i.e.}
with very high temperatures, contribute to the high energy tail of
the antiproton spectrum.
It should be noticed that the only reason
why they do not also dominate the low energy part is that the mass
spectrum behaves like
$M^2$ below $M_*$. In a Friedman universe without inflation, this
important feature would
make the accurate choice of the mass spectrum lower bound $M_{min}$
irrelevant (as long as it remains much smaller than $M_*$),
due to the very small number density of black
holes in this mass region.

\begin{figure}
\centerline{\includegraphics*[width=\textwidth]{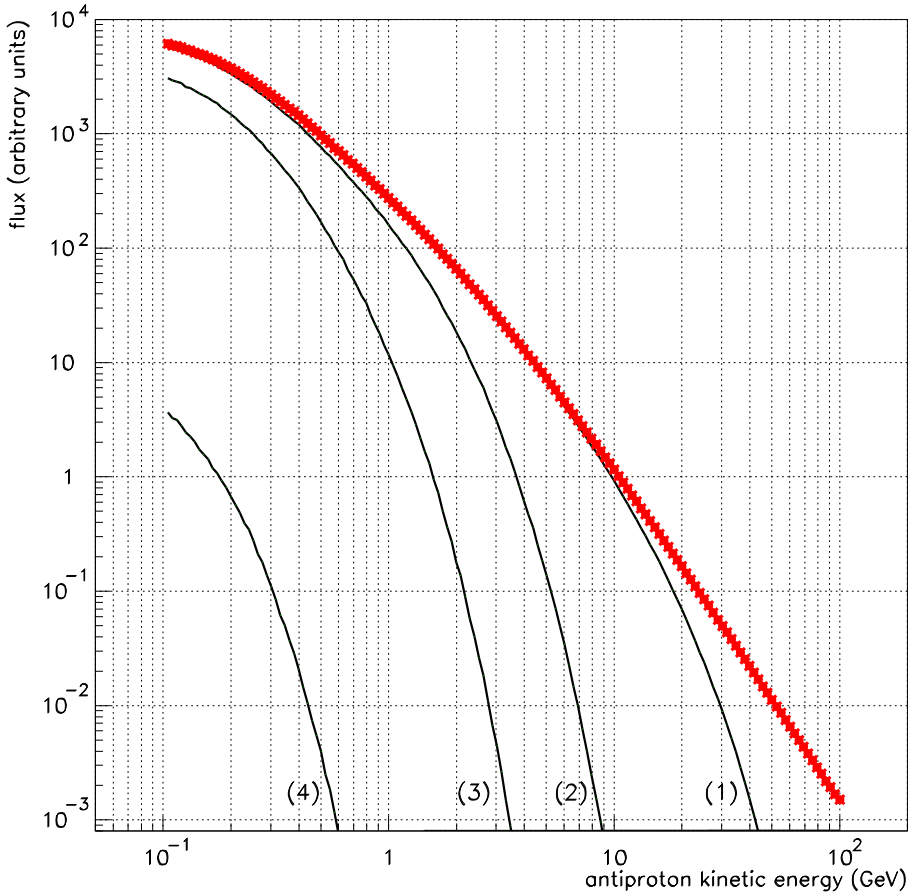}}
\caption{Primary antiproton flux with the standard mass spectrum
before propagation (in arbitrary
units) with different reheating temperatures. Curve (1) is for
$T_{RH}=3\cdot 10^9$ GeV, curve (2) is for
$T_{RH}=10^9$ GeV, curve (3) is for $T_{RH}=6\cdot 10^8$ GeV, curve (4)
is for $T_{RH}=3\cdot 10^8$ GeV and the thick line is the spectrum without
any cutoff.
}\label{reheating}
\end{figure}

Fig.~\ref{reheating} shows how inflation modifies the {\it standard}
mass spectrum. The
reheating temperature used herafter is nothing else
than a way of taking into
account a cutoff in the mass spectrum due to the rather large horizon
size after inflation.
There is nearly no constraint, neither on the theoretical side nor
on the observational
one, on this reheating temperature. A reliable lower limit can only be
imposed by phenomenological arguments around the nucleosynthesis
values, {\it i.e.} in the MeV
range (Giudice {\it et al.} \cite{Giudice}). On the other hand, upper
limits, taking into
account the full spectrum of inflaton decay products in the
thermalization process, are
close to $10^{12}$ GeV (McDonald, \cite{McDonald}). The main point
for the antiproton
emission study is that there is clearly a critical reheating
temperature, $T_{RH}^c\approx
10^9$ GeV . If $T_{RH} > T_{RH}^c$, the {\it standard} mass spectrum
today is nearly not modified
by the finite horizon size after inflation,
whereas if $T_{RH} < T_{RH}^c$, the
minimal mass becomes so large
that the emission is strongly reduced.
It is important to notice, first, that a rather
high value of $T_{RH}^c$ makes the existence of {\sc pbh}s evaporating now
into antiprotons quite unlikely, second, that the flux is varying very
fast with $T_{RH}$ around $T_{RH}^c$.

\begin{figure}
\centerline{\includegraphics*[width=\textwidth]{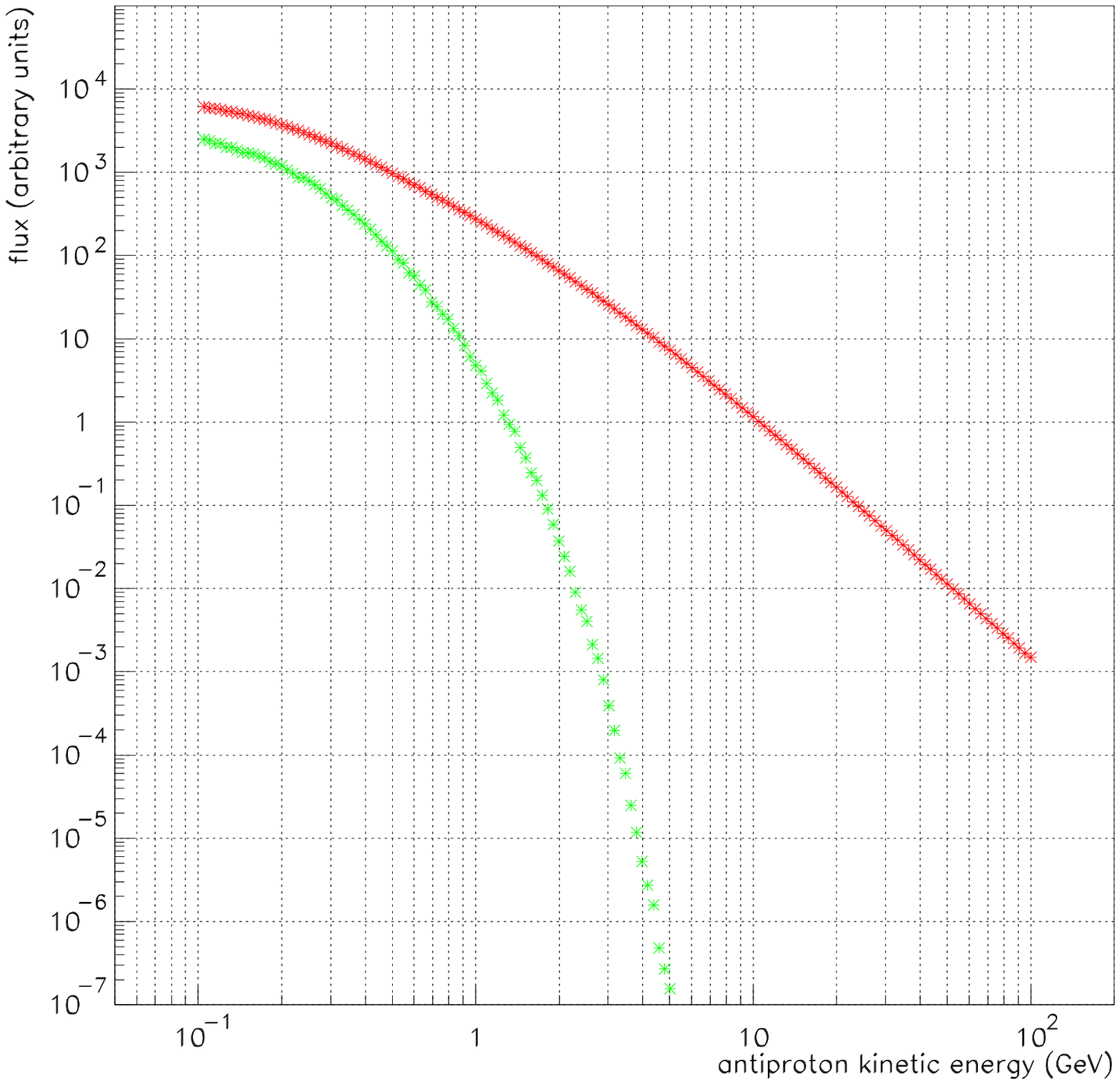}}
\caption{Primary antiproton flux with the standard mass spectrum. The
higher points correspond
to the pure Hawking spectrum and the lower points to the modified law
with a QCD halo.
}\label{halo}
\end{figure}

It could also be mentioned that above some critical temperature,
the emitted Hawking radiation could interact with itself and form a
nearly thermal
photosphere (Heckler \cite{Heckler}). This idea was numerically
studied (Cline {\it et al.} \cite{Cline}) with the full Boltzmann
equation for the particle
distribution and the Hawking law as
a boundary condition at horizon. For the antiproton emission, the
relevant process is the
formation of a quark-gluon plasma which induces energy losses before
hadronization. We have
taken into account this possible effect using the spectrum given by
Cline {\it et al.} (\cite{Cline}) : ${\rm d}^2N/{\rm d}E{\rm d}t
\propto \exp(-E/T_0)$
where $T_0 \approx 300$ MeV normalized to the accurately computed
Hawking spectrum.
The resulting effect, dramatically reducing the evaluated flux, is
shown in Fig.~\ref{halo}.

\section{Propagation model}
\label{propagation_model}
\subsection{Features of the model}
\label{propagation_model1}
The propagation of cosmic rays throughout the Galaxy is described
with a two--zone
effective diffusion model which has been thoroughly discussed
elsewhere (Maurin {\it et al.} 2001 - hereafter Paper I, Donato {\it et
al.} 2001a - hereafter Paper II).
We repeat here the main features of this diffusion model for the sake
of completeness but we refer the reader to the above-mentioned papers
for further details and justifications.

\begin{figure}
\centerline{\includegraphics*[width=\textwidth]{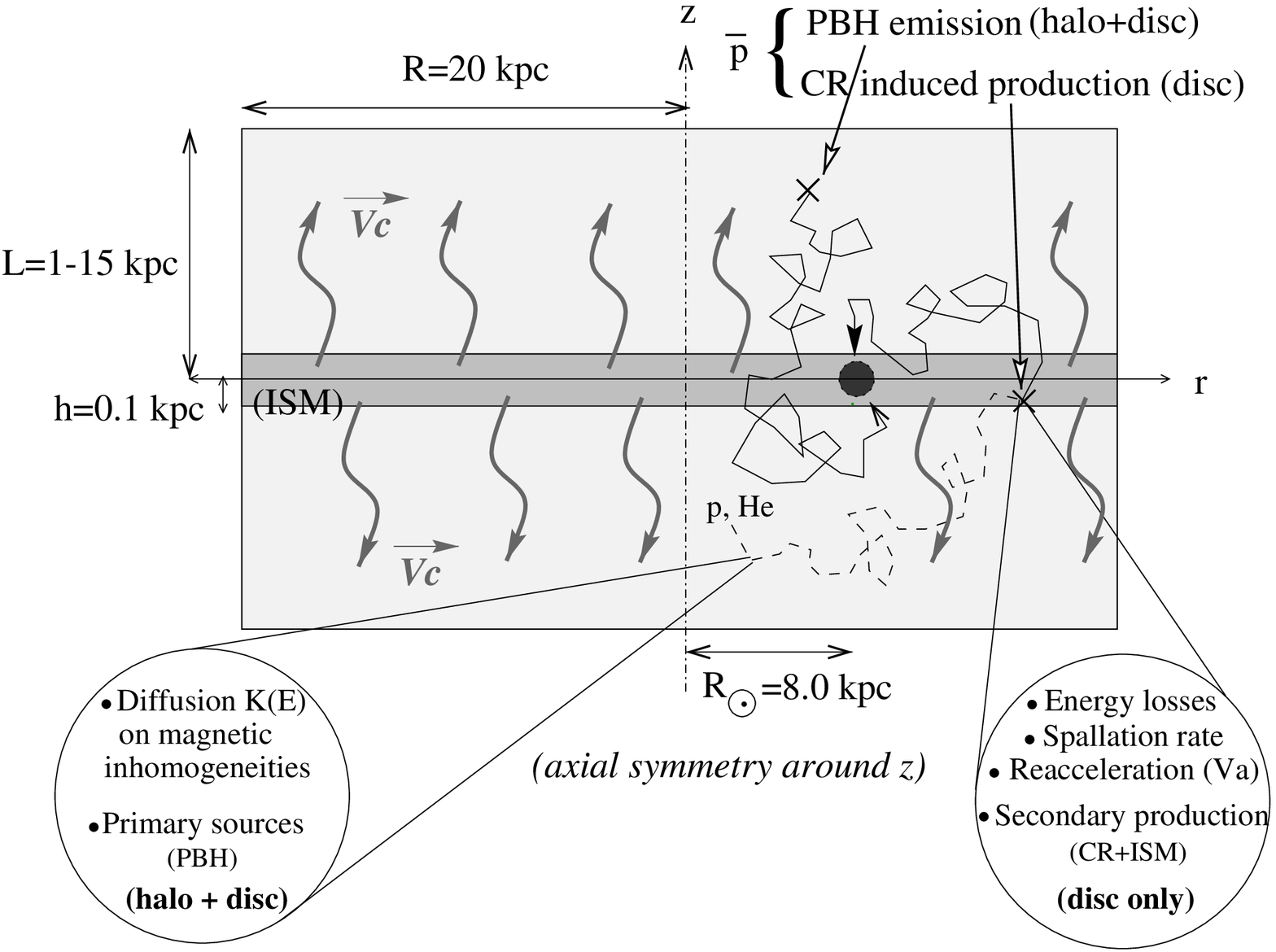}}
\caption{Schematic view of the axi-symmetric diffusion
model. Secondary antiproton sources originate from {\sc cr}/{\sc ism}
interaction in the disc only; primary sources
are also distributed in the dark halo which extends far beyond the diffusion
halo.
In the latter case, only sources embedded in the diffusion halo contribute
to the signal (see Appendix~\ref{AppB}).}
\label{diffusion}
\end{figure}

The Milky--Way is pictured
as a thin gaseous disc with radius $R = 20$ kpc and thickness
$2 h = 200$ pc (see figure~\ref{diffusion}) where charged nuclei
are accelerated and destroyed by collisions on the interstellar gas,
yielding secondary cosmic rays.
That thin ridge is sandwiched between two thick confinement layers of
height $L$, called {\em diffusion halo}.
The five parameters of this model are $K_0$, $\delta$,
describing the diffusion
coefficient $K(E) = K_0 \beta {\cal R}^{-\delta}$,  the halo
half-height $L$, the convective velocity $V_c$ and the Alfven
velocity $V_a$.
Specific treatment related to $\bar{p}$ interactions (elastic
scattering, inelastic destruction) and more details can also be found
in Paper II.

Actually, a confident range for these five parameters has been obtained
by the analysis of charged stable cosmic ray nuclei data (see Paper I).
The selected parameters have been employed in Paper II to
study the secondary antiproton flux, and are used again in this analysis
(for specific considerations about the Alfv\'en
velocity, see section 5.2 of Paper II).
In principle, this range could be further reduced using
more precise data or considering different sorts of
cosmic rays. For the particular case of $\beta$ radioactive nuclei,
Donato {\it et al.} \cite{Fiorenza3} showed that with existing data
no definitive and strict conclusions can so far be drawn.
We thus have chosen a conservative
attitude and  we do not apply any cut  in our initial sets of parameters
(which can  be seen in figures 7 and 8 of Paper I).

\subsection{Solution of the diffusion equation for antiprotons}

Antiproton cosmic rays have been detected, and most of them were
probably secondaries, {\it i.e.} they were produced by nuclear reactions of a
proton or He cosmic ray {\sc (cr)} nucleus impinging on interstellar
{\sc (ism)}
hydrogen or helium atoms at rest.
When energetic losses and gains are discarded, the secondary
density $N^{\bar{p}}$ satisfies the relation (see Paper II for details)
\begin{eqnarray}
         2 \, h \, \delta(z) \, q^{sec}(r,0,E) & = &
         2 \, h \, \delta(z) \, \Gamma^{ine}_{\bar{p}} \, N^{\bar{p}}(r,0,E)
         \nonumber \\
         & + &
         \left\{
         V_{c} \frac{\partial}{\partial z} \, - \,
         K \left(
         {\displaystyle \frac{\partial^{2}}{\partial z^{2}}} +
         \frac{1}{r} {\displaystyle \frac{\partial}{\partial r}}
         \left( r {\displaystyle \frac{\partial}{\partial r}} \right)
         \right) \right\} \, N^{\bar{p}}(r,z,E) \;\; ,
         \label{EQUATION_GENERALE}
\end{eqnarray}
as long as steady state holds.
Due to the cylindrical geometry of the problem, it is easier to extract
solutions performing Bessel expansions of all quantities
over the orthogonal set of Bessel functions
$J_{0}(\zeta_{i} x)$ ($\zeta_{i}$ stands
for the ith zero of $J_{0}$ and $i = 1 \dots \infty$).
The solution of equation~(\ref{EQUATION_GENERALE}) may be written as
(see eqs A.3 and A.4 in paper II)
\begin{equation}
N^{\bar{p},\;sec}_{i}(z,E) \; = \; {\displaystyle \frac{2 \,
h}{A_{i}}} \;
q_{i}^{sec}(E) \;\times
\exp \left\{ {\displaystyle \frac{V_{c} \, |z|}{2 \, K}} \right\} \;
\left\{
{\sinh \left\{ {\displaystyle \frac{S_{i}}{2}} \, \left( L - |z| \right)
\right\}} \, / \,
{\sinh \left\{ {\displaystyle \frac{S_{i}}{2}} \, L \right\}} \right\}
\;\; ,
\label{solution_sec}
\end{equation}
where the quantities $S_{i}$ and $A_i$ are defined as
\begin{equation}
S_{i} \equiv \left\{
{\displaystyle \frac{V_{c}^{2}}{K^{2}}} \, + \,
4 {\displaystyle \frac{\zeta_{i}^{2}}{R^{2}}}
\right\}^{1/2}
\;\;\; \mbox{and} \;\;\;
A_{i}(E) \equiv 2 \, h \, \Gamma^{ine}_{\bar{p}}
\; + \; V_{c} \; + \; K \, S_{i} \,
{\rm coth} \left\{ {\displaystyle \frac{S_{i} L}{2}} \right\}
\;\; .
\label{definition_Ai}
\end{equation}

We now turn to the primary production by {\sc pbh}s. It is described by a
source term distributed over all the dark matter halo
(see section \ref{distrib})
-- this has not to be confused with the diffusion halo -- whose core
has a typical size of a few kpc.
At $z=0$ where fluxes are measured, the corresponding density is given
by (see Appendix~\ref{AppA})
\begin{equation}
         N^{\bar{p},prim}_{i}(0)= \exp\left( \frac{-V_cL}{2K} \right)
         \frac{y_i(L)}{A_i\sinh(S_iL/2)}
         \label{solprim1}
\end{equation}
where
\begin{eqnarray}
         y_i(L)= 2\int_0^L\exp\left( \frac{V_c}{2K}(L-z')\right)
         \sinh\left(\frac{S_i}{2}(L-z')\right)q^{prim}_{i}(z')dz'.
         \label{igrec}
\end{eqnarray}


This is not the final word, as the antiproton spectrum is affected by
energy
losses when $\bar{p}$ interact with the galactic interstellar matter
and by energy  gains when reacceleration occurs.
These energy changes are described by the integro--differential equation
\begin{equation}
         A_{i} \, N^{\bar{p}}_{i} \; + \; 2 \, h \, \partial_{E}
         \left\{ b^{\; \bar{p}}_{loss}(E) \, N^{\bar{p}}_{i} \, - \,
         K^{\; \bar{p}}_{EE}(E) \, \partial_{E} N^{\bar{p}}_{i} \right\}
          = 2 \, h \, \left\{
         q_{i}^{prim}(E) + q_{i}^{sec}(E) +
         q_{i}^{ter}(E) \right\}
         \label{bob}
\end{equation}
We added a source term $q_{i}^{ter}(E)$, leading to the
so-called tertiary component.
It corresponds to inelastic but non-annihilating reactions of $\bar{p}$
on interstellar matter, as discussed in Paper II.
The resolution of this equation proceeds as described in Appendix~(A.2),
(A.3) and Appendix (B) of Paper II, to which we refer for further
details.
The total antiproton flux is finally given by
\begin{equation}
         N^{\bar{p},\;tot}(R_{\odot},0,E) \; = \;
         {\displaystyle \sum_{i=1}^{\infty}} \,
\left(N^{\bar{p},\;sec}_{i}(0,E)
         +N^{\bar{p},\;prim}_{i}(0,E) \right)\,
         J_{0} \left( \zeta_{i} {\displaystyle \frac{R_{\odot}}{R}} \right)
\end{equation}
where $N^{\bar{p},\;sec}_{i}(0,E)$ and $N^{\bar{p},\;prim}_{i}(0,E)$
are given by formul{\ae}~(\ref{solution_sec}), (\ref{solprim1}) and
(\ref{igrec}).
We emphasize that the code (and thus numerical procedures) used in this
study is exactly the same as the one we used in our previous analysis
(Paper I and II), with the new primary source term described above.

As previously noticed, the dark halo extends far beyond the diffusion
halo whereas its core is grossly embedded within $L$. We can wonder if
the external sources not comprised in the diffusive halo significantly
contribute to the amount of $\bar{p}$ reaching Earth.
A  careful analysis shows that in the situation studied here, this
contribution can be safely neglected (see Appendix~\ref{AppB}).


\section{Top of atmosphere spectrum}

\subsection{Summary of the inputs}

For the numerical results presented here, we have considered the
following source term, which is a particular case of the profiles
discussed in section \ref{distrib}
\begin{equation}
       q^{prim}(r,z,E) = \rho^{PBH}_\odot
        \left( \frac{R_c^2 + R_\odot^2}{R_c^2 +  r^2 + z^2}
        \right) \times Q^{PBH}(E)
        \label{plotch}
\end{equation}
where $Q^{PBH}(E)$ was shown in figures (\ref{pbar}), (\ref{reheating})
and (\ref{halo}) for various assumptions.
It corresponds to the isothermal case with a core radius $R_c=3$ kpc
(the results are modified only by a few percents if $R_c$ is varied
between 2 and 6 kpc).
A possible flattening of this halo has been checked to be irrelevant.
Actually the Moore or NFW halos do not need to be directly computed for
this study as they only increase the total
{\sc pbh} density inside the diffusion volume for a given local density:
as a result the flux is higher and the more conservative upper limit is
given by the isothermal case.

The primary source term (eq. \ref{plotch}) is injected in eqs. (\ref{solprim1})
and (\ref{igrec}),
then added to the standard secondary contribution (eq. \ref{solution_sec}),
and propagated (eq. \ref{bob}) for a given set of parameters (see section
\ref{propagation_model1}).
Once this interstellar flux has been calculated, it must be corrected
for the effects of the solar wind, in order to be compared with the
top of atmosphere observations.
We have obtained the {\sc toa} fluxes following the usual force field
approximation (see Donato {\em et al.} \cite{Fiorenza2} and
references therein).
Since we will compare our predictions with data taken during a period of
minimal solar activity, we have fixed the solar modulation parameter
to $\phi = 500$ MV.

To summarize, the whole calculation is repeated for
many $Q^{PBH}(E)$ input spectra, and all possible values of the propagation
parameters $V_c$, $L$, $K_0$, $V_a$ $\delta$ before being modulated.
The modulated flux is then compared to data to give constraints
on the local {\sc pbh} density $\rho^{PBH}_\odot$.

\subsection{Features of the propagated {\sc pbh} spectrum}
\begin{figure}
\centerline{\includegraphics*[width=\textwidth]{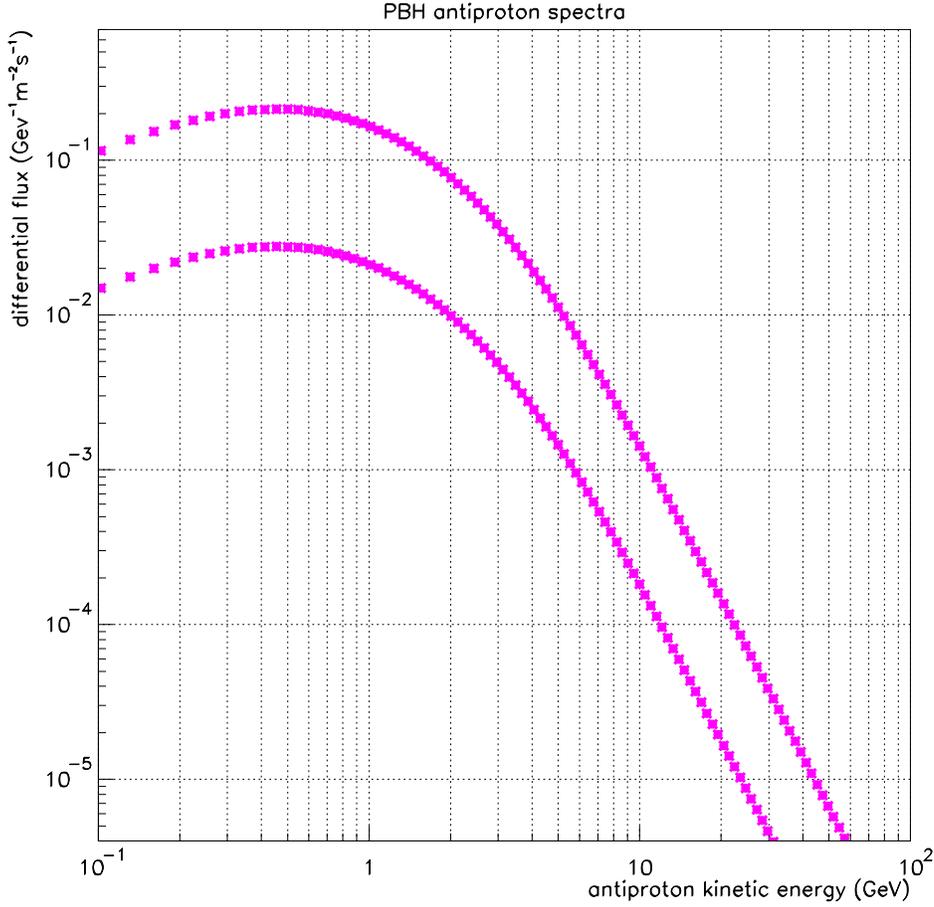}}
\caption{TOA primary antiproton flux with the standard mass spectrum
after propagation for
$\rho_{\odot}^{PBH}=10^{-32}~{\rm g}\, {\rm cm^{-3}}$ and extreme
astrophysical models.
}\label{pbar_primaire}
\end{figure}

Fig.~\ref{pbar_primaire} shows the top of atmosphere
antiproton flux resulting from the {\it standard} mass spectrum
without any inflation cutoff ($T_{RH}>T_{RH}^c$) for a local density
$\rho_{\odot}^{PBH}=10^{-32}~{\rm g}\, {\rm cm^{-3}}$. The two
curves correspond to the extreme astrophysical models considered as
acceptable in the extensive study of nuclei below $Z=30$ (paper I).
We can notice that uncertainties due to astrophysical parameters are very
important on the primary propagated $\bar{p}$ flux.
The degeneracy in the diffusion parameters that one can observe for stable
nuclei --  and also for secondary antiprotons (see figure 7 of Paper II)
-- is broken down for {\sc pbh} $\bar{p}$'s.

This can be easily understood: secondary $\bar{p}$ and {\sc cr} nuclei
are created by spallations in the galactic disc, so that all the
sets of diffusion parameters that make the primaries cross the same grammage
($\sim 20$ g cm$^{-2}$) give the same secondary flux.
The simplest diffusion models (no reacceleration, no galactic wind -- see
for example Webber {\em et al.} \cite{WebberLeeGupta}) predict that
the only relevant parameter is then $K_0/L$.
On the other hand, primary antiprotons sources are located in the dark
matter halo, and their flux is very sensitive to the total quantity
of sources contained in the diffusion halo, {\it i.e.} to $L$.
This explains the scatter of about one order of magnitude in predicted
fluxes that are shown in Fig.~\ref{pbar_primaire} for a given local density.

\subsection{Comparison with previous works}
In our models, all sets of parameters compatible with B/C data have 
been retained.
In particular, a wide range of values for $L$ is allowed ($L\geq 1$
kpc) and  each $L$ is
correlated with the other parameters.
The upper limit is set to 15 kpc, which is motivated by physical
arguments (see for example Beuermann {\it et al.} \cite{Beuermann}
and Han {\it et al.}
\cite{Han} by direct observation of radio halos in galaxies, showing
that $L$ is a few kpc, and Dogiel \cite{Dogiel} for a compilation
of evidences).

To our knowledge, the most complete studies about primary {\sc pbh} antiprotons
are those by Mitsui {\it et al.} (\cite{Mitsui}) and Maki {\it et al.}
(\cite{Orito}) in which the authors use roughly adjusted values for
diffusion parameters and restrict their analysis to two cases, $L=2$  kpc
and $L=4$ kpc (and only one value of $K_0/L = 0.008$~kpc~Myr$^{-1}$).
This is somewhat a crude estimation of what we know (or don't know) about $L$.
Moreover, in their models, neither galactic wind nor
reacceleration is considered. The second point should not be very
important (see Paper II) whereas the first point has a crucial impact on
the detected antiproton flux.
It can be shown that fluxes are
at least exponentially  decreased  by
the presence of a Galactic wind: setting $V_c=0$ may overestimate
the number of primary $\bar{p}$ reaching Earth.

Effects of propagation parameters on primaries are qualitatively discussed in
Bergstr\"om {\em et al.} (\cite{Bergstrom}) for the case of {\sc susy} primary
antiprotons.
At variance with all previous works on the subject, our treatment allows a
systematic and quantitative estimation of these uncertainties, as the
allowed range of
propagation parameters is known from complementary cosmic ray analysis.

The difference in the spectral properties we consider
are less crucial:
the input {\sc pbh} spectrum used here is mostly the same as in Maki 
{\it et al.}
(\cite{Orito}), whereas the accurate dimensionless absorption 
probability for the
emitted species is taken into account instead of its high energy limit and the
possible presence of a QCD halo is also investigated.

\section{Upper limit on the {\sc pbh} density}

\subsection{Method}

\begin{figure}
\centerline{\includegraphics*[width=\textwidth]{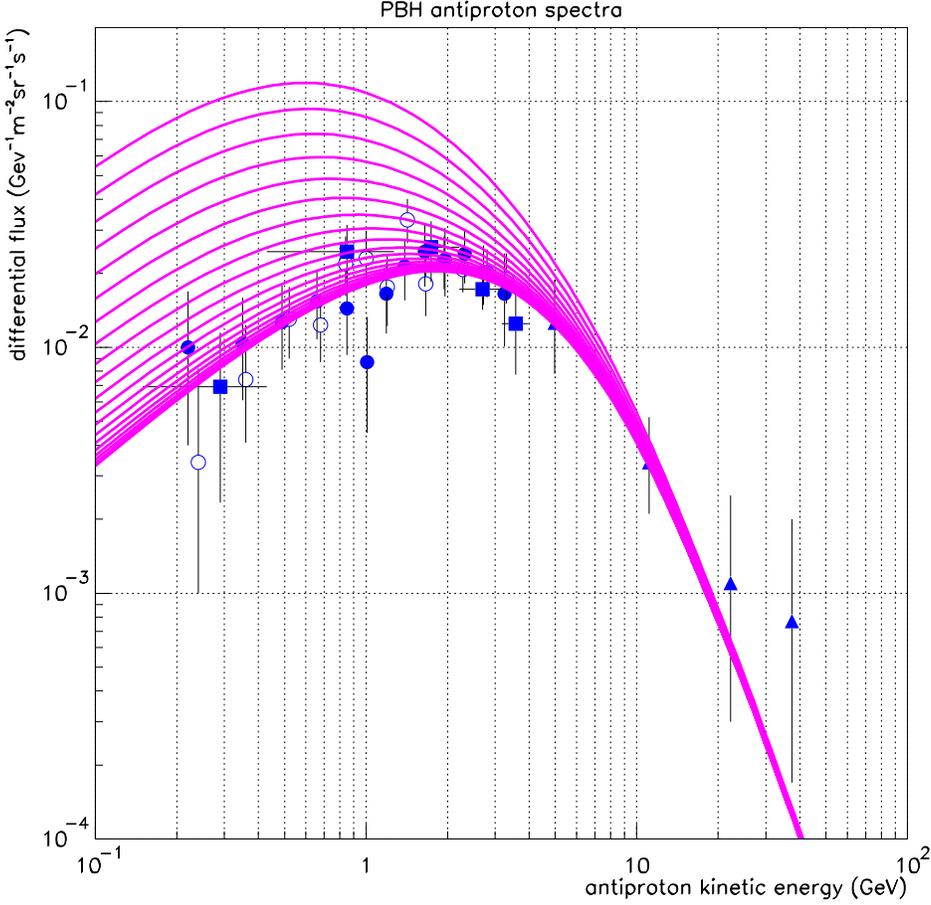}}
\caption{Experimental data from BESS95 (filled circles), BESS98 (circles),
CAPRICE (triangles) and AMS (squares) superimposed
with mean theoretical {\sc pbh} spectra for $\rho_{\odot}^{PBH}$ between
$5\cdot10^{-35}~{\rm
g}.{\rm cm}^{-3}$ (lower curve) and $10^{-32}~{\rm g} \,{\rm cm}^{-3}$
(upper curves).
}\label{pbar_tot}
\end{figure}

Of course, the previously mentioned uncertainties will clearly weaken
the usual upper limits on
{\sc pbh} density derived from antiprotons. To derive a reliable upper
limit, and to account for asymmetric
error bars in data, we define a generalized $\chi^2$ as

\begin{displaymath}
      \begin{array}{ll}
        \chi^2= &\sum_i
        \frac{(\Phi^{th}(Q_i)-\Phi_i^{exp})^2}
        {(\sigma^{exp+}_i+\sigma^{th+}(Q_i))^2}\Theta(\Phi^{th}(Q_i)-\Phi^{exp}_i)\\
        & +\sum_i \frac{(\Phi^{th}(Q_i)-\Phi_i^{exp})^2}
        {(\sigma^{exp-}_i+\sigma^{th-}(Q_i))^2}\Theta(\Phi^{exp}_i-\Phi^{th}(Q_i)).\\
      \end{array}
\end{displaymath}
where $\sigma^{th+}$ and $\sigma^{exp+}$ ($\sigma^{th-}$ and $\sigma^{exp-}$)
are the theoretical and experimental
positive (negative) uncertainties.
Superimposed with BESS, CAPRICE and AMS data (Orito {\it et al.} \cite{Orito1},
Maeno {\it et al.} \cite{Maeno}, CAPRICE Coll. \cite{CAPRICE}, Alcaraz {\it et
al.} \cite{Alcaraz}), the full antiproton flux,
including the secondary component and the primary component, is shown
on Fig.~\ref{pbar_tot} for 20 values of $\rho_{\odot}^{PBH}$
logarithmically spaced between $5 \, 10^{-35}$
and $10^{-32}~{\rm g} \,{\rm cm^{-3}}$.
The {\it standard} mass spectrum is assumed and one astrophysical model is
arbitrarily chosen, roughly corresponding to the average set of free
parameters.

An upper limit on
the primary flux for each value of the magnetic halo thickness is computed.
The theoretical errors included in the $\chi^2$ function
come from nuclear physics ($p+He \rightarrow \bar{p}+X$
   and $He+He \rightarrow \bar{p}+X$, Paper II) and from the 
astrophysical parameters analysis
which were added linearly, in order to remain conservative. The resulting
$\chi^2$ leads to very conservative results as it assumes that limits on the
parameters correspond to 1 sigma.

\subsection{Results}

\begin{figure}
\centerline{\includegraphics*[width=\textwidth]{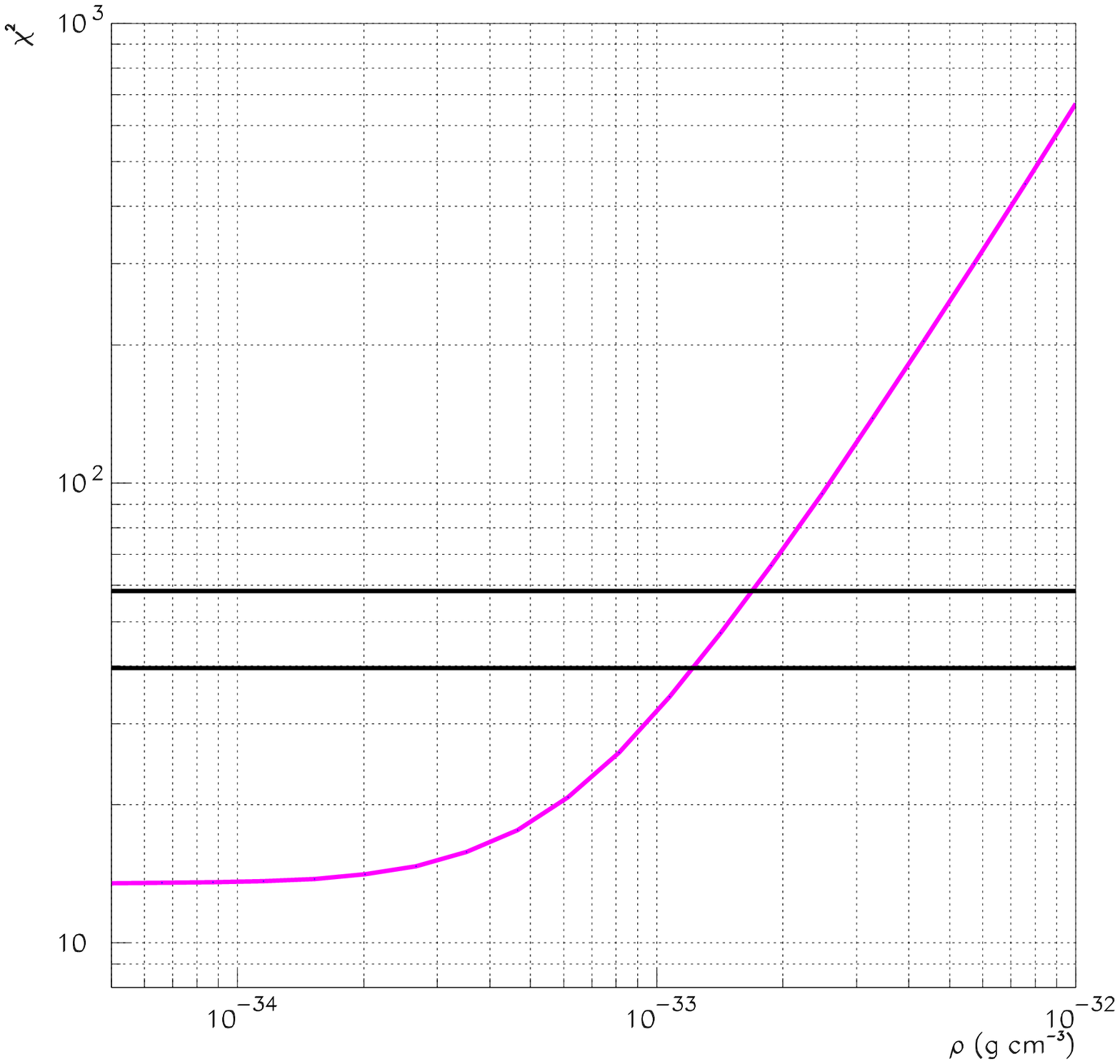}}
\caption{$\chi^2$ between the experimental data and the theoretical flux
as a function of the local density of {\sc pbh}s for a magnetic halo thickness
of 3 kpc. The horizontal lines are 99\% and 63\% confidence levels.
}\label{chi2look}
\end{figure}

Fig.~\ref{chi2look}
gives the $\chi^2$ as a function of $\rho_{\odot}^{PBH}$ for $L=3$ kpc.
The horizontal
lines correspond to 63\%
and 99\% confidence levels. In this paper the statistical significance
of such numbers should be taken with care : they only refer to orders
of magnitude.
As expected, the $\chi^2$ value is constant for small {\sc pbh}
densities (only secondaries contribute to the flux) and is
monotonically increasing without
any minimum : this shows that no {\sc pbh} (or any other primary source)
term is needed to
account for the observed antiproton flux.

\begin{figure}
\centerline{\includegraphics*[width=\textwidth]{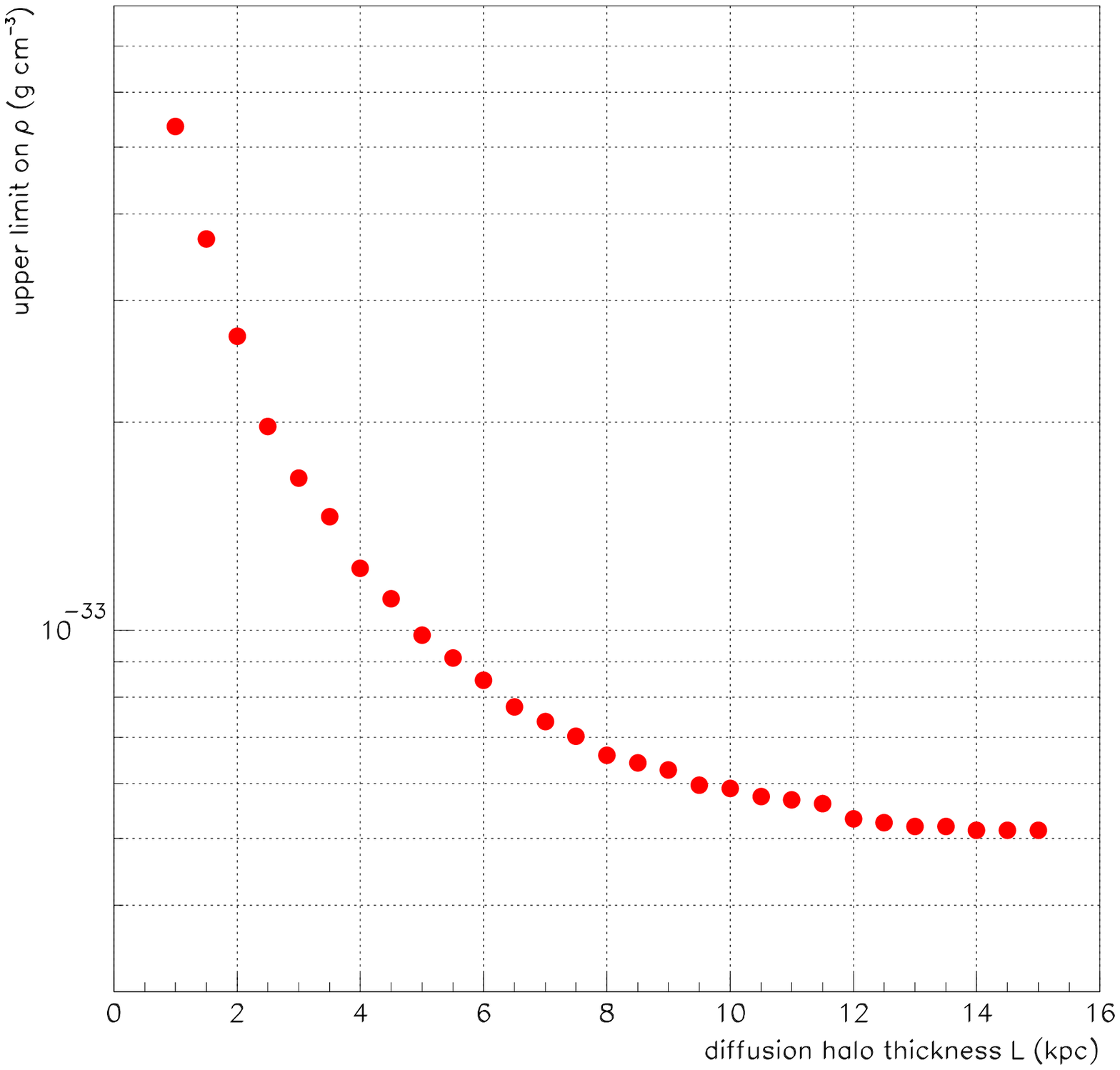}}
\caption{Upper limits on the local density of {\sc pbh}s as a function of
the magnetic halo thickness L.
}\label{chi2}
\end{figure}

Fig.~\ref{chi2} gives the upper limits on the local density of {\sc pbh}s as
a function of $L$ with the {\it standard} mass spectrum. It is a
decreasing
function of the halo size because a bigger diffusion region means a higher
number of {\sc pbh}s inside the magnetic zone for a given local density.
Between $L=1$ and $L=15$ kpc (extreme astrophysical values), the 99\%
confidence
level upper limit goes from $5.3\cdot 10^{-33}~{\rm g} \,{\rm cm}^{-3}$ to
$5.1\cdot 10^{-34}~{\rm g} \,{\rm cm}^{-3}$.

An important point shown in section (\ref{cumul}) is
that we are sensitive only to
{\sc pbh}s with masses between $10^{12}$g and $10^{14}$g. The upper limit
given here on the total local mass density of {\sc pbh}s can therefore
be given in a ``safer" way as a numerical density of {\sc pbh}s
integrated between
$10^{12}$g and $10^{14}$g. Although less interesting from the point of
view of cosmology, this value has the great interest of being
independent of the initial mass spectrum shape
as this remains within the ``increasing" part of the mass
spectrum.
The resulting numerical density is
$n_{\odot~PBH}=3.9\cdot 10^{-51}~{\rm
cm}^{-3}$ for $L=3$ kpc.

If we take into account the possible QCD photosphere (see figure \ref{halo})
around {\sc pbh}s, the previous upper limit is substantially
weakened:
$\rho_{\odot}^{PBH} < 1.2 \cdot 10^{-32}~{\rm g}\, {\rm cm^{-3}}
$ for the "usual" $L=3$ kpc.
In this
case, gamma-rays are a much more
powerful tool to study {\sc pbh}s. It should, nevertheless, be
emphasized that this model is still controversial. First of all,
because  the quarks and gluons which are
below threshold for pion production seem to be simply ignored, then,
because the assumption that the
Brehmsstrahlung interaction has a range of $1/m_e$ could be wrong for
hard interactions, which would
lead to a great overestimation of the photosphere effect (Cline,
private communication).

\section{Discussion and future prospects}

\subsection{Comparison with other existing limits}

Many upper limits on the {\sc pbh} explosion rate have been derived
thanks to 100 MeV, 1 TeV and 100 TeV $\gamma$-rays or low-energy antiprotons.

In the ultra-high energy range, a reliable search for short gamma-ray bursts
radiation from an arbitrary direction have been performed using the
CYGNUS air-shower array (Alexandreas {\it et al.} \cite{Alex}).  No strong
one second burst was observed and the
resulting upper limit is
$dN_{\odot exp}^{PBH}/dt\, d^3V \leq
0.9\cdot10^6~{\rm year}^{-1}{\rm pc}^{-3}$.
Very similar results were derived by the Tibet (Amenomori {\it et
al.} \cite{Ame}) and the AIROBIC
collaborations (Funk {\it et al.} \cite{Funk}).
TeV gamma-rays have also been used to search for short
time-scale coincidence events. The bursts detected
are compatible with the expected background and the resulting upper
limit obtained with 5 years of data
(Connaughton \cite{Connaughton})  is $dN_{\odot exp}^{PBH}/dt\, d^3V\leq
3\cdot10^6~{\rm year}^{-1}{\rm pc}^{-3}$.

The limit coming from antiprotons has been advocated to be far better:
the previous study from Maki {\it et al.} (\cite{Orito}) gives
$dN_{\odot exp}^{PBH}/dt\, d^3V \leq 2\cdot10^{-2}~{\rm year}^{-1}{\rm
pc}^{-3}$.
However, we emphasize that this limit does not take into account the 
wide range of
possible astrophysical uncertainties (in particular $L$, which can
affect the limits by one order of magnitude).
Moreover, we believe that the explosion rate is not
the pertinent variable to use when comparing
results from different approaches: thresholds differences
between experiments make the meaning of "explosion" very different.
With a mass spectrum $\propto M^2$ for
small masses, the number of exploding {\sc pbh}s depends strongly on
the value of the threshold. It makes the comparison of our results with
the ones from Maki {\it et al.} very ambiguous.
This is why we prefer to give our upper limit, either as
a local mass density, assuming a {\it standard} mass spectrum,
$\rho^{PBH}_{\odot} < 5.3\cdot 10^{-33}~{\rm g} \,{\rm cm}^{-3}$,
either as a number density, independant of the mass spectrum shape outside the
relevant interval,
$n_{\odot}^{PBH} < 1.3 \cdot 10^{-50}~{\rm cm}^{-3}$ (whatever L).

Gamma-rays in the 100 MeV region provide a sensitive 
probe to the
presence of PBHs along the line of sight up to redshifts as large
as $\sim 700$. Gamma-rays in this energy range have little interactions with the
intergalactic medium and can travel on cosmological distances. The
integration of the signal involves therefore a much larger scale than
in the case of Milky Way antiprotons. It should also be pointed out that
in this case most of the PBH population is involved as the dominant 
emission peaks above
100 MeV even at the present epoch. By matching the PBH cosmological
emission to the extra-galactic gamma-ray diffuse background
(MacGibbon \& Carr \cite{MacGibbon2}), the limit
$\Omega_{PBH} \leq 1.8 \times 10^{-8} \, h^{-2}$ was obtained.
Being mostly based on the assumption that a standard PBH mass 
spectrum holds above $M_*$, this result is robust.
Our limit ($\rho^{PBH}_{\odot} < 5.3 \times 10^{-33}~{\rm g} \,{\rm cm}^{-3}$)
also requires the same assumption. It no longer depends on
the details of cosmic-ray propagation as it corresponds to the
minimal possible value of 1 kpc for $L$. This
result is therefore quite conservative. In order to discuss it in the light
of the gamma-ray constraints, it should be noticed that any PBH population
is a particular form of cold dark matter. When the latter collapses
to form galactic halos and the intra-cluster medium, PBHs merely
follow the collapse. Their abundance in the solar neighbourhood
should trace their cosmological contribution to the overall value
of $\Omega_M \sim 0.3 - 0.4$. Because a canonical isothermal halo has
a solar density of $\sim 0.3$ GeV cm$^{-3}$
-- $\sim 5.3 \times 10^{-25}$ g cm$^{-3}$ --
we infer an upper limit of $\sim 10^{-8}$ on the contribution of PBHs
to the galactic -- and as mentionned above to the cosmological -- dark
matter. Our antiproton bound translates into
$\Omega_{PBH} \leq 10^{-8} \, \Omega_M \sim 4 \times 10^{-9}$.
Such a constraint is comparable to the limit derived from gamma-ray
considerations. Antiprotons are produced by PBHs that are exploding
at the present epoch and the limit which they provide is complementary
to the gamma-ray constraint.

\subsection{Possible improvements on the antiproton limit}

New data on stable nuclei along with a better understanding of diffusion
models (see for example Donato et al., in preparation) could allow
refinement in propagation to constrain $L$.
It is also important to notice that the colour emission treatment could
probably be improved.
Studying more accurately the colour field confinement and the effects of
angular momentum quantization,
it has been shown (Golubkov {\it et al.} \cite{Golubkov}) that the meson
emission is modified. Those ideas have not yet been applied to
baryonic evaporation.

Several improvements could be expected for the detection of
antimatter from {\sc pbh}s in the years to come.
First, the AMS experiment (Barrau \cite{Barrau3}) will allow, between
2005 and 2008, an extremely precise measurement
of the antiproton spectrum.
In the meanwhile, the BESS (Orito {\it al.} \cite{Orito2}) and
PAMELA (Straulino {\it et al.} \cite{PAMELA}) experiments should have
gathered new high-quality data.
The solar modulation effect should be taken
into account more precisely to discriminate between primary and
secondary antiprotons as the shape of each component will not be affected in
the same way. The effect of polarity should also be included (Asaoka
{\it et al.}
\cite{Asaoka}).

To conclude, the antideuteron signal should also be studied as
it could be the key-point to distinguish between
{\sc pbh}-induced and {\sc susy}-induced antimatter in cosmic rays.
Although the $\bar{p}$ emission due to the annihilation of
supersymetric dark matter would have nearly the same spectral characteristics
than the {\sc pbh} evaporation signal, the antideuteron production 
should be very
different as coalescence
schemes usually considered (Chardonnet {\it et al.} \cite{Chardonnet}) cannot
take place between successive {\sc pbh} jets.

\begin{acknowledgements}
The authors would like to thank J.H. Mc Gibbon for very helpful
discussions and D. Page for providing the absorption cross-sections.
\end{acknowledgements}



\appendix

\section{Computation of the primary galactic flux}
\label{AppA}
The solution for a generic source term $q^{prim}(r,z,E)$,
is obtained by making the substitution
\begin{equation}
         2 \, h \, \delta(z) \, q^{sec}(r,0,E) \;\;\;
         \longleftrightarrow \;\;\;
         q^{prim}(r,z,E)
\end{equation}
in equation~(\ref{EQUATION_GENERALE}).
The Bessel expansion of this term now reads
\begin{equation}
         q^{prim}(r,z,E) \; = \;
         {\displaystyle \sum_{i=1}^{\infty}} \, q_{i}^{prim}(z,E) \,
         J_{0} \left( \zeta_{i} {\displaystyle \frac{r}{R}} \right) \;\; ,
\end{equation}
with (introducing $\rho\equiv r/R$)
\begin{equation}
         q_{i}^{prim}(z,E) =\frac{2}{J^{2}_{1}(\zeta_{i})}
         \int_{0}^{1}\rho\;
         q^{prim}(\rho,z,E)J_{0}(\zeta_{i}\rho)\;d\rho
\end{equation}
The procedure to find the solutions of  equation~(\ref{EQUATION_GENERALE})
is standard (see Paper I, Paper II and references therein).
After Bessel expansion, the equation to be solved in the halo is
\begin{equation}
         K\left[ \frac{d^{2}}{dz^{2}}-\frac{V_{c}}{K}
         \frac{d}{dz}- \frac{\zeta_{i}^{2}}{R^{2}}\right]
         N^{\bar{p},prim}_{i}(z)
         =- q^{prim}_{i}(z)
\end{equation}
Using the boundary condition $N^{\bar{p}}_i(z=L)=0$, and
assuming continuity between disc and halo, we obtain
\begin{eqnarray}
      N^{\bar{p},prim}_{i}(z)&=& \exp\left( \frac{V_c(|z|-L)}{2K}\right)
      \frac{y_i(L)}{A_i\sinh(S_iL/2)} \nonumber \\
      &&\left[
\cosh(S_iz/2)+\frac{(V_c+2h\Gamma^{ine}_{\bar{p}})}{KS_iA_i}\sinh(S_iz/2)\right]
      -\frac{y_i(z)}{KS_i}
\end{eqnarray}
where
\begin{eqnarray}
         y_i(z)= 2\int_0^z\exp\left( \frac{V_c}{2K}(z-z')\right)
         \sinh\left(\frac{S_i}{2}(z-z')\right)q^{prim}_{i}(z')dz'
\end{eqnarray}
In particular, at $z=0$ where fluxes are measured, we have
\begin{equation}
     N^{\bar{p},prim}_{i}(0)= \exp\left( \frac{-V_cL}{2K} \right)
     \frac{y_i(L)}{A_i\sinh(S_iL/2)}.
     \label{eq_finale}
\end{equation}

\section{Sources located outside the diffusive halo}
\label{AppB}
\begin{figure}[h!]
      \centerline{\includegraphics*[width=0.55\textwidth]{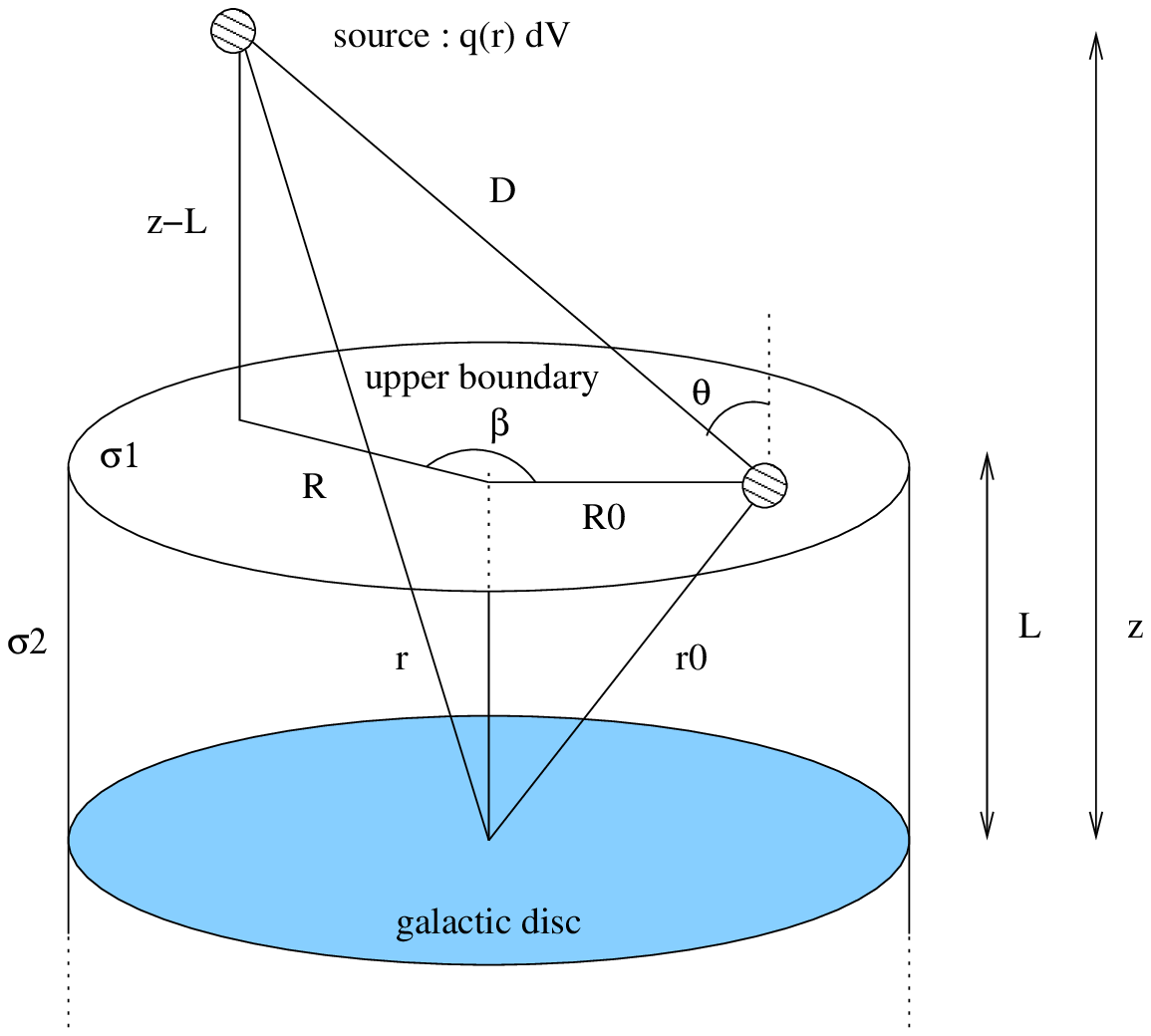}}
      \label{notations_bords}
\end{figure}

By definition, diffusion is much less efficient outside of the diffusive halo,
so that antiprotons coming from these outer sources propagate almost freely,
until they reach the diffusion box boundary.
At this point, they start to interact with the diffusive
medium and after travelling a distance of the order of the mean free
path, their propagation becomes diffusive.
Thus, the flux of incoming antiprotons gives rise to three thin
sources located at the surfaces $\sigma_1$, $\sigma_2$ and $\sigma_3$
(see figure \ref{notations_bords}),
the first with a $z$ distribution $\eta(z)$ which has non-zero
values only for $z \sim L$ (the upper boundary of the diffusive
volume), the second with a $r$ distribution $\eta(r)$ which has non-zero
values only for $r \sim R$ (the side boundary) and the third located
at $z \approx -L$ (the lower boundary).

\subsection{Solar density of antiprotons coming from external
sources}

Let us focus first on the top and bottom surfaces $\sigma_1$ and
$\sigma_3$ located at $z \sim L$.
As expression (\ref{solprim1}) was obtained for a symmetrical source
term, the corresponding external source term can be directly written as
\begin{displaymath}
        q^{ext}(r,z) = {\cal F}^{inc}(r) \eta(L-z)
        \;\;\; \mbox{with} \;\;\;
        \int \eta(z) dz = 1
\end{displaymath}
where $\eta(z) dz$ is the probability for an incoming nucleus to interact
with the diffusive medium in the layer comprised between distances $z$ and
$z+dz$ from the boundary,
and ${\cal F}^{inc}(r)$ is the antiproton flux coming from outside
the diffusive halo.
The function $\eta(z)$ is expressed in kpc$^{-1}$.
Using the fact that the surface source is located around small values of
$u \equiv L-z$, insertion of the above relation in formula (\ref{igrec}) gives
\begin{displaymath}
        y_i^{ext}(L) \approx 2 {\cal F}^{inc}_i
        \frac{S_i}{2}
        \int u \eta(u)du
        = {\cal F}^{inc}_i(r) S_i \bar{u}
\end{displaymath}
where ${\cal F}^{inc}_i$ is the Bessel transform of ${\cal
F}^{inc}(r)$ and where the mean free path
$\bar{u}=\int u  \eta(u)du$
has been introduced, so that (\ref{solprim1}) becomes
\begin{equation}
         N^{ext}_{i}(0)= \exp\left( \frac{-V_cL}{2K}\right)
         \frac{{\cal F}^{inc}_i S_i \bar{u}}{A_i\sinh(S_iL/2)}.
\end{equation}
Before turning to the computation of the incoming flux
${\cal F}^{inc}(r)$ at the box boundary, we can make a remark about the
above expression. If galactic wind and spallations are neglected, it
can be simplified into
\begin{displaymath}
      N^{ext}_{i}(0) \approx
      \frac{{\cal F}^{inc}_i \bar{u}}{K\cosh(\zeta_iL/R)}.
\end{displaymath}
It occurs that the diffusion coefficient $K$ and the mean free path
$\bar{u}$ are related in a way that depends on the microscopic
details of the diffusion process. For hard spheres diffusion,
$\bar{u}= 3K/v$ so that
\begin{displaymath}
      N^{ext}_{i}(0) \approx \frac{3{\cal F}^{inc}_i}{v\cosh(\zeta_iL/R)}.
\end{displaymath}
The physical interpretation of this expression is interesting: for
low Bessel indices $i$, {\color{red} {\it i.e.}} for large spatial scales,
the $\cosh(\zeta_iL/R)$
term is close to unity and the Bessel terms in the disk density are
proportional
to those in the source term.
For large Bessel indices $i$, {\it i.e.} for small spatial
scales, the Bessel terms are exponentially lowered in the disk density compared
to the source term.
The small scale variations of the source are filtered out by the
diffusion process. This filtering is more efficient for large values of
the diffusion zone height $L$.

In the following, we show the results for the full expression
\begin{equation}
      N^{ext}_i(0)= \exp\left( \frac{-V_cL}{2K} \right)
      \frac{{\cal F}^{inc}_i S_i 3K}{\beta c A_i\sinh(S_iL/2)}
      \label{penible}
\end{equation}
where the hard sphere expression has been used for $\bar{u}$.

\subsection{Computation of the incoming flux ${\cal F}^{inc}(r)$ from
external sources}

The flux of nuclei reaching the surface from a source
located outside the diffusive halo is
\begin{displaymath}
      d{\cal F}^{inc} = \cos \theta \;
      \frac{q_{PBH}(\vec{r}) \, dV}{4\pi D^2}.
\end{displaymath}
The integration over the angular coordinate $\beta$ is performed
using the property (see Gradshteyn \cite{Gradshteyn})
\begin{displaymath}
        \int_0^\pi \frac{d\beta}{(a-b \cos \beta)^{3/2}}
        = \frac{2 {\mathbf E}(x)}{(a-b) \sqrt{a+b}}
\end{displaymath}
where ${\mathbf E}$ is the elliptic function (denoted $R_F$ in Numerical
Recipes, \cite{nr}) and where $x \equiv \sqrt{2b/(a+b)}$.
Finally, we can write
\begin{displaymath}
        {\cal F}^{inc}(r) = \int_{R=0}^\infty
\int_{z=L}^\infty
        \frac{ q_{PBH}(\vec{r})}{4\pi}
        \frac{2 {\mathbf E}(x) \, (z-L)\, R \,  dR \,
dz}{\left((z-L)^2+(R-R_0)^2\right)
        \sqrt{(z-L)^2+(R+R_0)^2}}.
\end{displaymath}
This quantity is then numerically computed and Bessel-transformed to
be incorporated in eq. (\ref{penible}).

\subsection{Conclusion: the external contribution is negligible}

The contribution of these external sources is shown in table
\ref{contribution_ext} for $L=1 \, \mbox{kpc}$ (the
lowest value we allowed) and $L=5 \, \mbox{kpc}$, with or without
galactic wind. It is larger for lower halo heights $L$, but it
is always less than $10^{-4}$ and it can be  safely neglected.
\begin{table}[ht]
        \caption{Fraction $N_{ext}/N_{tot}$ of the antiproton density in the
        solar neighbourhood due to the external primary sources, for the {\sc
        pbh} density profiles discussed in the text. The relevant
        cross-sections
        for antiproton spallations have been considered in the $A_i$
        term, and a kinetic energy of 1 GeV has been assumed. The
        results are not very sensitive to this particular
        energy value.}
        \label{contribution_ext}
        \begin{center}
          \begin{tabular}{|l|cc|cc|}
              \hline
              & \multicolumn{2}{c|}{$V_c = 0\,\mbox{km/s}$} &
              \multicolumn{2}{c|}{$V_c = 10\,\mbox{km/s}$}\\ \cline{2-5}
              density profile & $L=1\,\mbox{kpc}$&
              $L=5\,\mbox{kpc}$ & $L=1\,\mbox{kpc}$ & $L=5\,\mbox{kpc}$\\
              \hline
              Moore & $1.7 \times 10^{-4}$ & $4.9 \times 10^{-6}$ &
              $4.9 \times 10^{-6}$ & $4.0 \times 10^{-15}$\\
              NFW & $1.8 \times 10^{-4}$ & $5.4 \times 10^{-6}$
              & $5.4 \times 10^{-6}$ & $4.3 \times 10^{-15}$\\
              Modified isothermal & $2.5 \times 10^{-4}$ & $9.3 \times
              10^{-6}$
              & $1.5 \times 10^{-5}$ & $7.0 \times 10^{-15}$ \\
              Isothermal ($a=1 \;\mbox{kpc}$) & $1.7 \times 10^{-4}$ &
              $4.3 \times 10^{-6}$
              & $1.0 \times 10^{-5}$ & $3.6 \times 10^{-15}$ \\
              \hline
          \end{tabular}
        \end{center}
\end{table}

Similar results apply to the side boundary  $\sigma_2$ which is about
$R_{gal} -
R_\odot \sim 12$ kpc away.
Roughly, the contribution is the same order of magnitude than would be obtained
with a thin source $\sigma_1$ located at $z=12$ kpc, which can
be neglected as discussed above.

Note also that we present here only two combinations of $V_c$ and
$L$ to derive the surface contribution, but all the good sets
of propagation parameters used throughout this paper (and also in Paper I
and Paper II) point toward the same values, which are not larger than those
derived in table 1. We can thus conclude that primary surface
contributions are negligible in all cases.

\end{document}